\documentclass{aastex63}
\usepackage{amsmath, bm}
\usepackage{color}
\accepted{by ApJ}
\shorttitle{Large-scale magnetic field advected in the corona}
\shortauthors{Li \& Cao}
\graphicspath{{./}}

\begin{document}

\title{The large-scale magnetic field advected in the corona of a thin accretion disk}

\correspondingauthor{Jia-wen Li \& Xinwu Cao}
\email{jwlizju@zju.edu.cn, xwcao@zju.edu.cn}

\author{Jia-wen Li}
\affiliation{Zhejiang Institute of Modern Physics, Department of Physics, Zhejiang University, \\
38 Zheda Road, Hangzhou 310027, China; jwlizju@zju.edu.cn, xwcao@zju.edu.cn}

\author{Xinwu Cao}
\affiliation{Zhejiang Institute of Modern Physics, Department of Physics, Zhejiang University, \\
38 Zheda Road, Hangzhou 310027, China; jwlizju@zju.edu.cn, xwcao@zju.edu.cn}
\affiliation{Shanghai Astronomical Observatory, Chinese Academy of Sciences, 80 Nandan Road, Shanghai, 200030, China}
\affiliation{Key Laboratory of Radio Astronomy, Chinese Academy of Sciences, 210008 Nanjing, China}




\begin{abstract}

Large-scale magnetic field is believed to play a key role in launching and collimating jets/outflows. It was found that advection of external field by a geometrically thin disk is rather inefficient, while the external weak field may be dragged inwards by fast radially moving tenuous or/and hot gas above the thin disk. We investigate the field advection in a thin (cold) accretion disk covered with hot corona, in which turbulence is responsible for the angular momentum transfer of the gas in the disk and corona. The radial velocity of the gas in the corona is significantly higher than that in the thin disk. Our calculations show that the external magnetic flux is eﬀiciently transported inwards by the corona, and the field line is strongly inclined towards the disk surface, which help launching outflows. The field configurations are consistent with those observed in the numerical simulations. The strength of the field is substantially enhanced in the inner region of the disk (usually several orders of magnitude higher than the external field strength), which is able to drive a fraction of gas in the corona into outflows. This mechanism may be useful in explaining the observational features in X-ray binaries and active galactic nuclei. Our results may help understanding the physics of the magneto-hydrodynamic (MHD) simulations.
\end{abstract}

\keywords{accretion, accretion disks --- 
galaxies: jets ---ISM: jets and outflows --- magnetic fields}

\section{Introduction} \label{sec:intro}
Jets and outflows are ubiquitous phenomenons in a wide range of accreting systems, and the large-scale magnetic field plays an important role in accelerating and collimating jets and/or outflows \citep[see reviews of][and the references therein]{2007prpl.conf..277P, 2010LNP...794..233S, 2019ARA&A..57..467B}. Ultra-fast outflows (UFOs) have been observed in many active galactic nuclei (AGNs)  \citep[e.g.,][]{2009ApJ...701..493R,2010A&A...521A..57T,2013MNRAS.430.1102T,2015Natur.519..436T,2015MNRAS.451.4169G,2017MNRAS.469.1553P,2020ApJ...895...37R}. The origin and acceleration mechanisms of such powerful outflows are still unclear. A large fraction of these UFOs are highly ionized, and therefore the UV or soft X-ray opacity is very small \citep[][]{2013MNRAS.430...60G,2011ApJ...742...44T}, which lead to inefficient line force acceleration \citep[e.g.,][]{2014ApJ...789...19H}. One of the most promising models for outflow acceleration is the Blandford-Payne mechanism \citep[][]{1982MNRAS.199..883B}, in which the kinetic power of a magnetized accretion disk is extracted by the large-scale magnetic field to accelerate the outflows. The relativistic jets observed in AGNs or black hole X-ray binaries may probably be driven by co-rotating large-scale magnetic field dragged by the rotating black hole \citep[][]{1977MNRAS.179..433B}. Strong large-scale magnetic field near the black hole is a necessary ingredient in the Blandford-Znajek mechanism. {Observations of powerful disk wind in both X-ray binaries and  AGNs may suggest the presence of significant poloidal magnetic flux threading the accretion disk \citep[][]{2006Natur.441..953M,2004MNRAS.355.1105F,2015ApJ...814...87M,2015ApJ...805...17F,2016ApJ...821..104Y,2018ApJ...853...40F, 2018ApJ...852...35K,2020ApJ...904...30M,2021ApJ...906L...2K}, since the large-scale magnetic field is expected to be responsible for driving these powerful outflows. The origin of such large-scale magnetic field, however, is still quite unclear. } 
 
It has been suggested that external weak magnetic field can be advected inwards by the accretion disk, which is fed by a companion star or the interstellar medium, provided they are somewhat magnetized   \citep[][]{1974Ap&SS..28...45B,1976Ap&SS..42..401B,1989ASSL..156...99V,2005ApJ...629..960S,2019Natur.573...83Z}. However, transport of the magnetic flux in a turbulent thin disk ($H/R \ll 1$) is too inefficient to attain sufficient magnetic flux, which is unable to drive outflows from the inner region of the disk \citep[][]{1989ASSL..156...99V,1994MNRAS.267..235L}, because the magnetic diffusivity is roughly proportional to turbulent viscosity, i.e., the Prandtl number is around unity \citep[][]{1979cmft.book.....P,2009A&A...507...19F,2009ApJ...697.1901G}.  

One of the attempts proposed to solve this issue is that the radial velocity of a thin disk will be substantially increased, if most of the angular momentum of the accreting gas is removed by the magnetically driven outflow, and then it makes the field advection much more efficient than a conventional thin disk without outflows  \citep[][]{2013ApJ...765..149C,2014ApJ...786....6L,2019ApJ...872..149L}. Another possibility is to assume highly conducting nonturbulent or fast  moving layers above a thin disk to reduce outward diffusion of the magnetic flux, which can enhance field advection in the disk \citep[][]{2009ApJ...701..885L,2012MNRAS.424.2097G,2013MNRAS.430..822G}. The local analyses of the field advection in the disk with vertically extended sphere have been carried out by some workers \citep[][]{2009ApJ...701..885L,2012MNRAS.424.2097G,2013MNRAS.430..822G}, which indeed show that the fast moving tenuous gas helps field advection. Another candidate of such fast moving gas may be the hot corona above the disk, which is most likely to emit power-law hard X-ray photons due to the inverse Compton scattering of the soft photons from the disk \citep[][]{1979ApJ...229..318G,1991ApJ...380L..51H,1993ApJ...413..507H,2009MNRAS.394..207C,2012ApJ...761..109Y}.  
Recently, some numerical simulations show that the magnetic flux is preferentially transported by the corona above the disk, and a quasi-static large-scale magnetic field with field lines strongly inclined towards the disk surface is formed, which could driven hot gas into outflows \citep[e.g.,][]{2018ApJ...857...34Z,2020MNRAS.492.1855M}.

In this work, we investigate the global magnetic field structure of a thin disk with fast moving gas (either tenuous gas or hot corona). Section \ref{sec:model} contains the model of field advection in a thin disk with fast moving gas.  
The numerical method is described in Section \ref{sec:num_set}. The results and discussion of the model calculations are given in Sections \ref{sec:result} and \ref{sec:discussion} respectively. The last section contains a summary of this work.

\section{Model} \label{sec:model}
We study the large-scale field advection in a turbulent thin accretion disk covered by a layer of tenuous gas or a hot corona. In principle, a large-scale magnetic field threading a rotating disk may drive outflows from the disk surface under certain circumstance \citep*[][]{1994A&A...287...80C}. Such outflows may carry away a fraction of the angular momentum of the disk, which leads to a complicated disk-outflow connection  \citep*[e.g.,][]{2006A&A...447..813F,2013ApJ...765..149C,2019ApJ...872..149L}. For simplicity, we have not included the outflows in our model. It means that our calculations are good approximations for the case of weak magnetic outflows driven from the disk, or they are lower limits of the field advected by the disk, because the field advection would be enhanced in the presence of outflows \citep[][]{2013ApJ...765..149C,2019ApJ...872..149L}. 
We consider two different cases of fast moving gas above the thin disk. Case 1: The gas is isothermal vertically, while the radial velocity is described by changing the values of viscosity parameter as a parameterized function of $z$ to mimic the analytical results of \citet{2012MNRAS.424.2097G,2013MNRAS.430..822G}; Case 2: The gas temperature increases with $z$ to mimic a hot corona above the disk.      

\subsection{Thin disk covered by fast moving gas}

The structure of a thin accretion disk ($H/R \ll 1$) can be derived analytically with vertically averaged disk equations for $\alpha$-viscosity \citep[][]{1973A&A....24..337S}. In this work, we will have to consider the disk with vertically extended fast moving gas. In cylindrical coordinates, the $R\varphi$-component of the shear stress tensor can be prescribed as

\begin{equation}\label{eq:T_Rphi}
     t _{R\varphi}=\rho \nu R \frac{d \Omega}{d R}=-\alpha p,
\end{equation}
where $\nu$ is the effective turbulent viscosity, $\alpha$ is the viscosity parameter, $p=\rho c_{\rm s}^2$ is the gas pressure, $\rho$ is the gas density, and $c_{\rm s}$ is the isothermal sound speed. The turbulent viscosity is

\begin{equation}\label{eq:nu}
    \nu(z)= \frac{2}{3}\frac{\alpha c_{\rm s}^2(z)}{\Omega_{\rm k}},
\end{equation}
where the approximations $\Omega \sim \Omega_{\rm k}$ and $d\Omega/d R\sim -3\Omega_{\rm k}/2R$ are adopted.  
For a steady accretion disk, its radial velocity is

\begin{equation}\label{eq:vR}
    v_{R}(R,z) = -\frac{3\nu(z)}{2R}=-\frac{\alpha c_{\rm s}^2(z)}{R \Omega_{\rm k}}.
\end{equation}

Hydrostatic equilibrium in the vertical direction gives 

\begin{equation}
    \frac{1}{\rho (z)}\frac{d }{d z} p(z)= -z \Omega_{\rm k}^2,
\end{equation}
which is a reasonable approximation in the absence of outflows or for weak outflows. It reduces to 

\begin{equation}\label{eq:hydro_equi}
    \frac{1}{\rho(z)}\frac{d }{d z}\rho(z) = -\frac{1}{c_{\rm s}^2(z)}\frac{d }{d z}c_{\rm s}^2(z) - \frac{\Omega_{\rm k}^2 z}{c_{\rm s}^2(z)},
\end{equation}
where $p=\rho c_{\rm s}^2$ is used.

\begin{figure}[h!]
\centering
\includegraphics[width=0.5\columnwidth]{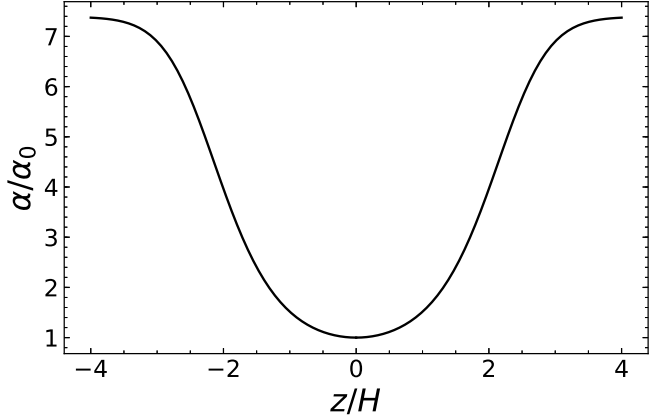}
\caption{Vertical profile of the $\alpha$ parameter, which is the same as that adopted in \cite{2013MNRAS.430..822G}  within the gas pressure dominate region (see the profile between the dotted lines in left-hand panel of Figure 1 in \cite{2013MNRAS.430..822G}). 
\label{fig:alpha}
}
\end{figure}

As discussed in \citet{2013MNRAS.430..822G}, the fast moving gas above the disk can be described with a vertically varying $\alpha(z)$, in which the gas is assumed to be isothermal vertically. 
Thus, the vertical structure of such a disk covered by tenuous gas can still be described by the conventional scale-height $H \equiv {c_{\rm s}}/{\Omega _{\rm k}}$ with $\Omega_{\rm k}$ is the Keplerian angular velocity, but the case of gas extending over several times $H$ is assumed in their calculations \citep[see][for the details]{2012MNRAS.424.2097G,2013ApJ...767...30B}. We choose the same $\alpha$ profile (see Figure \ref{fig:alpha}) as that used in their work, i.e., 

\begin{equation}
 \alpha (z) = \alpha_0 \frac{e^{2}}{1.0 + \left(e^{2}  - 1 \right)e^{-0.5\frac{z^2}{H^2}}  }
\end{equation}
where $\alpha_0$ is the value at disk midplane.

The radial velocity of the gas is then given by 

\begin{equation}\label{eq:vR_thinDisk}
    v_R(R,z) = -\frac{3\nu(z)}{2R} =  -\frac{3\alpha (z) c_s H}{2R}.
\end{equation}

The disk-corona system is usually referred to as two-phase model, since it can be described with two temperatures vertically for the cold disk and the hot corona respectively \citep[][]{1993ApJ...413..507H}. The temperature of the gas changes sharply from the cold disk to the corona in the vertical direction. we choose the following function to describe the gas temperature $\Theta(z)$ along $z$ direction,

\begin{equation}\label{eq:Theta_z}
   \Theta(z) = \left( \Theta_{z_{\rm h}} - \Theta_{0} \right) \frac{e^{b\left( 1 - a \right)} +1}{e^{b\left( 1 - a\frac{z}{z_{\rm h}} \right)} +1} + \Theta_{0},
\end{equation}
where the dimensionless gas temperature is defined as

\begin{equation}\label{eq:def_Theta}
    \Theta\equiv \frac{c_{\rm s}^2}{R^2 \Omega_{\rm k}^2},
\end{equation}
$\Theta_{0}$ and $\Theta_{z_{\rm h}}$ are the dimensionless gas temperature at the disk mid-plane, and the surface of the corona (i.e., $z=z_{\rm h}$), respectively. The location of the disk transiting to the corona, and the corona thickness are described by two parameters $a$ and $b$. Although an artificial function (\ref{eq:Theta_z}) is employed to describe the vertical temperature distribution of the disk-corona system, we believe it indeed reflects the basic feature of two-phase disk-corona model.        
 
Integrating Equation \eqref{eq:hydro_equi} along vertical direction from $z=0$ to $z=z_{\rm h}$, we obtain the vertical structure of the disk-corona system at radius of $R$, 
 
\begin{equation}\label{eq:int_hydro_equi}
    {\rm ln}\frac{\rho(z_{\rm h})}{\rho_0} = -{\rm ln} \frac{\Theta_{z_{\rm h}}}{\Theta_{0}} - \int_{0}^{z_{\rm h}} \frac{z}{R^2 \Theta(z)}d z.
\end{equation}

Finally, we need to determine the upper surface of the corona, and the gas density decreases with $z$ from disk mid-plane. The corona surface (i.e., $z=z_{\rm h}$, hereafter we refer $z=\pm z_{\rm h}$ as the corona surfaces) is defined as the gas density being a critical value $\epsilon\rho _0$, i.e.,

\begin{equation}\label{eq:epsison}
    \frac{\rho(z_{\rm h})}{\rho _0}=\epsilon,
\end{equation}
where $\rho_0$ is the gas density at the disk mid-plane. In this work, we focus on the field advection in the disk corona system, and therefore the gas pressure should be less than the magnetic pressure above the corona surface. Thus, the value of $\epsilon$ can be derived, provided the magnetic field strength at the corona surface is known.

Combining Equations \eqref{eq:Theta_z}, \eqref{eq:int_hydro_equi} and \eqref{eq:epsison}, we obtain the vertical structure of the disk-corona system [i.e., $z_{\rm h}$, and $\rho(z)$], when the parameters, $a$, $b$, $\Theta_{0}$, and $\Theta_{z_{\rm h}}$ are specified.

\subsection{Large-scale magnetic field advected in the disk with fast moving gas}
The advection of magnetic field $\boldsymbol{B}$ in a turbulent disk is described by induction equation, 

\begin{equation}\label{eq:induc_B}
\frac{\partial \boldsymbol{B}}{\partial t}=\nabla \times(\boldsymbol{v} \times \boldsymbol{B}-\eta \nabla \times \boldsymbol{B}),
\end{equation}
where $\boldsymbol{v}$ is the velocity of the plasma and $\eta$ is the magnetic diffusivity. In cylindrical coordinates, we assume an axisymmetric large-scale poloidal magnetic field threading an accretion disk, which can be described by a stream function $\psi(R,z)$ \citep[][]{1994MNRAS.267..235L}, i.e., $ \boldsymbol{B}_{\rm p} = \nabla \times \left[\psi(R,z)\boldsymbol{e}_{\varphi} \right]$, thus

\begin{equation}\label{eq:br_bz}
\begin{aligned}
B_{R} =& -\frac{\partial \psi}{\partial z},\\
B_z =& \frac{1}{R} \frac{\partial}{\partial R}\left( R \psi \right).
\end{aligned}
\end{equation}

Equation \eqref{eq:induc_B} can be rewritten as

\begin{equation}\label{induc_psi}
\frac{\partial }{\partial t} \left[ R \psi(R,z) \right] = -v_{R}(R,z)
\frac{\partial }{\partial R}
\left[ R \psi(R,z) \right] - \frac{4 \pi \eta}{c}R J_ \phi(R,z),
\end{equation}
where $v_{R}$ and $J_\varphi$ are the radial velocity of the gas and the azimuthal current density of the disk, respectively. The azimuthal current density $J_ \varphi (R,z)$ is related to the stream function $\psi(R,z)$ via the Biot-Savart law,

\begin{equation}\label{psi_d}
    \psi_{\rm d} (R,z)
    =\frac{1}{c} \int_{R_{\rm in}}^{R_{\rm out}} \int_{0}^{2\pi} \int_{-H}^{H} \frac{J_\varphi (R',z') \cos \varphi' d\varphi' R' d R' dz'}{\left[ R^2 + R'^{2} + \left( z-z'\right)^{2} - 2R R' \cos \varphi'\right]^{\frac{1}{2}}} ,
\end{equation}
where $c$ is the light speed and $\psi_{\rm d} (R,z)$ is contributed by the current $J_ \varphi (R',z')$ inside the disk (including the tenuous gas or the corona). We assume an external uniform vertical magnetic field $B_{\rm ext}$ to be advected inwards by the disk, and therefore

\begin{equation}\label{eq:psi_infty}
    \psi _\infty = \frac{1}{2}B_{\rm ext}R.
\end{equation}

The large-scale field advected by the disk is now described by the stream function $ \psi(R,z) = \psi_{\rm d}(R,z)+\psi_\infty$. The stream function contributed by the currents inside the disk can be rewritten as \citep[][]{jackson1999classical}

\begin{equation} \label{eq:psi_KE}
    \psi_{\rm d} (R,z) = \frac{1}{c}\int_{R_{\rm in}}^{R_{\rm out}} \int_{-H}^{H} \frac{4 J_\varphi(R',z')}{\left[ \left( R+R'\right)^2 +\left( z-z'\right)^2 \right]^{\frac{1}{2}}}\left[ \frac{\left(2-{\rm k}\right)K(\rm k) - 2 \emph{E} ({\rm k})}{\rm k} \right]R' dR' dz',
\end{equation}
where $\rm k$ is defined by 

\begin{equation}
    {\rm k} = \frac{4RR'}{\left( R+R' \right)^2 + \left( z-z' \right)^2},
\end{equation}
$K({\rm k})$ and $E({\rm k})$ are the complete elliptic integrals  

\begin{equation}
    K\left({\rm k}\right) = \int_{0}^{\frac{\pi}{2}} \left( 1 - {\rm k} \sin^2 \theta \right)^{-\frac{1}{2}}d \theta, 
\end{equation}
and

\begin{equation}
    E\left({\rm k}\right) = \int_{0}^{\frac{\pi}{2}} \left( 1 - {\rm k} \sin^2 \theta \right)^{\frac{1}{2}}d \theta.
\end{equation}

For a steady disk, i.e., $\partial / \partial t = 0$, the field advection is in balance with the outward diffuse of the magnetic flux, which leads to

\begin{equation}\label{patial_psi_d}
 -\frac{\partial}{\partial R} \left[ R \psi_{\rm d} (R,z) \right] - \frac{4\pi \eta}{c} \frac{R}{v_R(R,z)} J_ \varphi (R,z) = B_{\rm ext} R,
\end{equation}
where $\eta$, the magnetic diffusivity, is related to the magnetic Prandtl number by

\begin{equation}\label{eq:pm_0}
    {\cal P}_{\rm m,0} = \eta/\nu_0,
\end{equation}
and $\nu_0$ is the effective turbulent viscosity at the disk mid-plane (i.e., $z=0$). 
Differentiating Equation
\eqref{psi_d} and then substituting it into Equation \eqref{patial_psi_d}, we obtain a set of linear equations

\begin{equation}\label{eq:linear}
    -\sum_{ j=1}^{\rm n} \sum_{ l=1}^{\rm m} J_\varphi (R_{ j},z_{l}) P_{i,j,k,l} \Delta R_j \Delta z_l
    -\frac{4\pi \eta}{c} \frac{R_i}{v_R(R_i,z_k)} J_ \varphi (R_i,z_k) = B_{{\rm ext},k} R_i,
\end{equation}
where 

\begin{equation*}
P_{i,j,k,l}= {\frac{1}{c}}\int_{0}^{2\pi}{\frac {\left[ R_j^2 + \left( z_k - z_l \right)^2 - R_i R_j {\cos}\varphi'\right]R_j}{\left[ {R_i}^2+{R_j}^2 + \left( z_k - z_l \right)^2 -2R_i{R_j}\cos{\varphi '}\right] ^{\frac{3}{2}}}}\cos \varphi'{d}\varphi '.
\end{equation*}
 
The subscripts $_{i,j,k,l}$ are labeled for the variables at radius $R_i,R_j$, and vertical position $z_k,z_l$. $J_\varphi(R_j,z_l)$ is the current density at position $(R_j, z_l)$, and $B_{{\rm ext},k} R_i = B_{\rm ext} R_i$ ($B_{\rm ext}$ is the strength of the external uniform vertical magnetic field). 

Solving a set of the linear equations (\ref{eq:linear}) with given disk structure, i.e., the radial velocity and the disk thickness, We can obtain the distribution of the current density within the disk, and then the spatial distribution of the magnetic stream function $\psi (R,z)$. The large-scale poloidal magnetic field is then calculated with Equation \eqref{eq:br_bz}. 

\section{numerical setup}\label{sec:num_set}

For the vertically isothermal case, the disk structure is described by the parameters $H$, and the vertical distribution of $\alpha$. The large-scale field advection can be calculated by solving the linear equation \eqref{eq:linear} with the given $\alpha$ profile (see Figure \ref{fig:alpha}), when the values of the parameters ${\cal P}_{\rm m,0}$, and $H/R$ are specified. 

For the disk-corona system, the linear equation \eqref{eq:linear} can be solved, when the values of the model parameters ${\cal P}_{\rm m,0} $, $\Theta_{0}$, $\Theta_{z_{\rm h}}$ and $\epsilon$ are specified. In order to describe a sharp temperature increase from the disk to corona, we adopt $a=3$, and $b=15$ in Equation \eqref{eq:Theta_z} for the calculations (see Figure \ref{fig:DC_Theta}).
 
The current density $J_\varphi$ is discretized on the center of the cell area $\Delta R_j \times \Delta z_l$. In radial direction we set $n$ grid cells distributed logarithmically between $R_{\rm in}$ and $R_{\rm out}$, while the $m$ grid cells in $z$-direction is adopted between $z=z_{\rm h}$ and $z=-z_{\rm h}$. In all of the calculation we adopt $n=100$ and $m=50$, which can achieve a good performance on accuracy. After solving a set of linear equations (i.e., Equation \eqref{eq:linear}), the spacial distribution of the magnetic potential can be calculated similarly to \cite{1994MNRAS.267..235L}. Equation \eqref{eq:psi_KE} can be written in a matrix form and then the magnetic potential contributed by the currents inside the disk is
 
\begin{equation}
    \left(R \psi_d \right)_{i,k} = \sum_{j=1}^{n}\sum_{l=1}^{m}Q_{i,j,k,l} J_\varphi(R_j, z_l) \Delta R_j \Delta z_l,
\end{equation}
where $ \left(R \psi_d \right)_{i,k}$ is the magnetic potential at $(R_i, z_k)$ contributed by all current within the disk, and $Q$ is a matrix defined by Equation \eqref{eq:psi_KE},

\begin{equation}
    Q_{i,j,k,l} = \frac{1}{c}\frac{4 R_i R_j}{\left[ \left( R_i+R_j\right)^2 +\left( z_k-z_l\right)^2 \right]^{\frac{1}{2}}}\left[ \frac{\left(2-{\rm k}\right)K(\rm k) - 2 \emph{E} ({\rm k})}{\rm k} \right],
\end{equation}
and 

\begin{equation}
    {\rm k} = \frac{4R_i R_j}{\left( R_i+R_j \right)^2 + \left( z_k-z_l \right)^2}.
\end{equation}
Evaluating the elements of matrix $Q$ at grid cells of $R_i=R_j$ and $z_k=z_l$ would leads to singular, we adopt a smooth method similar to that used in \cite{1994MNRAS.267..235L}:

\begin{equation}
    Q_{j,j,l,l} = \frac{\lambda}{4}\left[Q_{j,j+\frac{\lambda}{2},l,l}+ Q_{j,j+\lambda,l,l}  + Q_{j,j-\frac{\lambda}{2},l,l}  + Q_{j,j-\lambda,l,l}
    + Q_{j,j,l,l+\frac{\lambda}{2}} + Q_{j,j,l,l+\lambda} +Q_{j,j,l,l -\frac{\lambda}{2}}+ Q_{j,j,l,l-\lambda}
    \right].
\end{equation}

Here, for grid cell $(R_j,z_l)$, $R_{j-\lambda}$ and $R_{j+\lambda}$ are the inner and outer boundaries, and  $z_{l+\lambda}$,  $z_{l-\lambda}$ are the upper and bottom boundaries, respectively, and we adopt $\lambda = 1/2$ in all the calculations. The final results are almost independent of the exact value of $\lambda$.

\section{results}\label{sec:result}

\begin{figure}
\centering
\includegraphics[width=0.5\columnwidth]{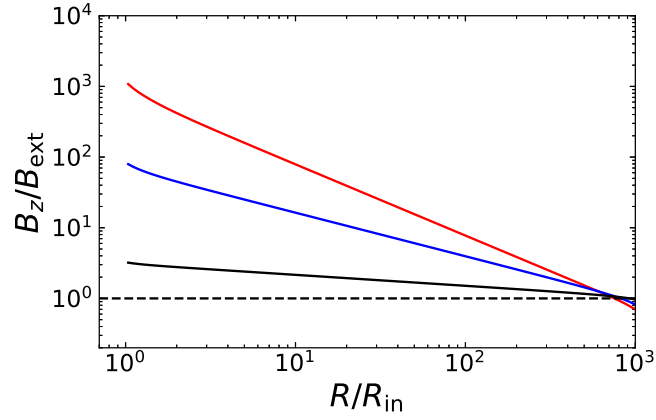}
\caption{Variations in the vertical component of the poloidal magnetic field at disk mid-plane with radius.  The red and blue lines show the results for $H/R=0.1$ and $H/R=0.05$, respectively. The black line is for a traditional turbulent thin disk with $H/R=0.1$, $B_{\rm ext}$ is the strength of the external imposed magnetic field (see Equation \eqref{eq:psi_infty}).   \label{fig:Bz_og}}
\end{figure}

\begin{figure}
\centering
\gridline{\fig{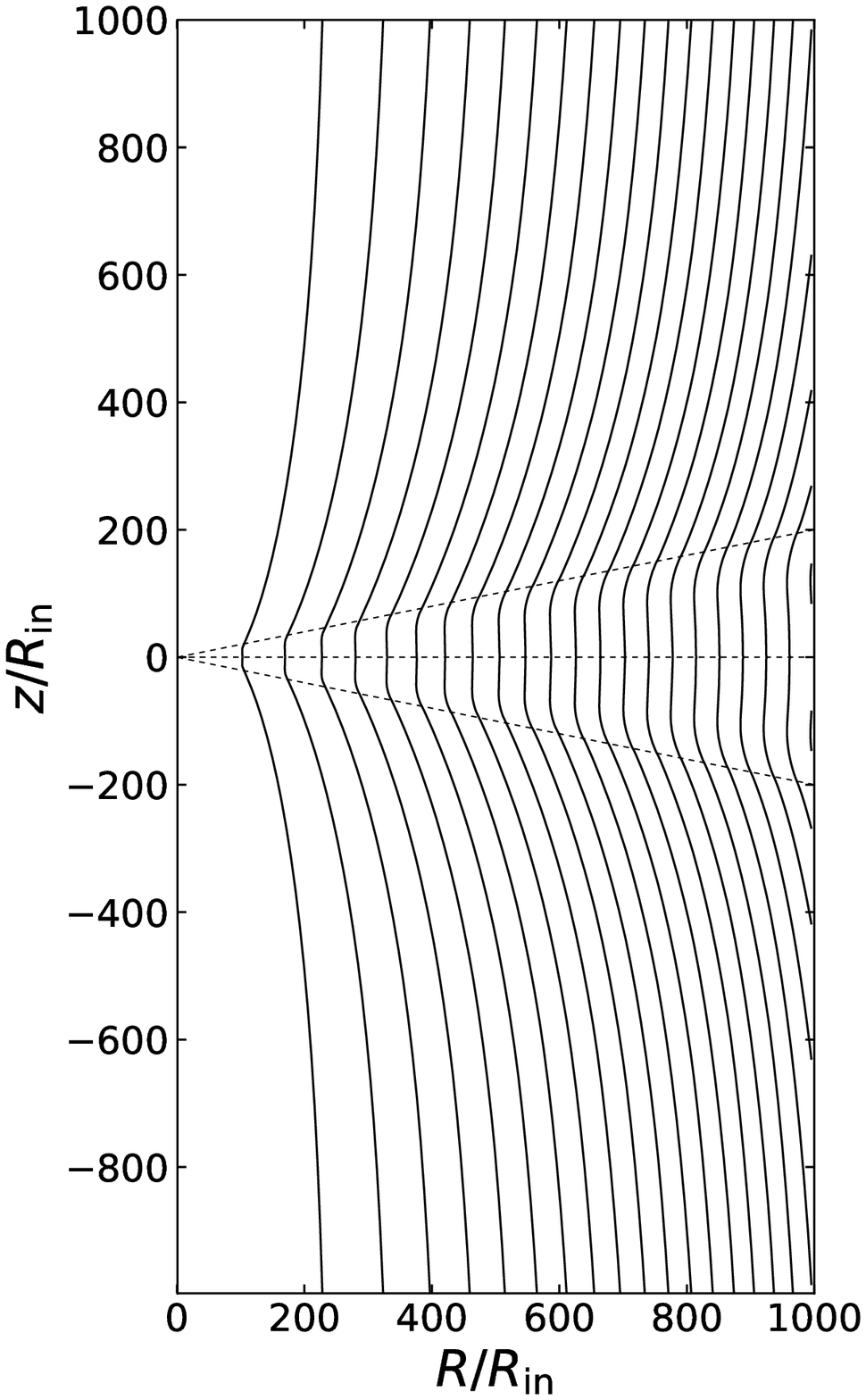}{0.48\textwidth}{}
          \fig{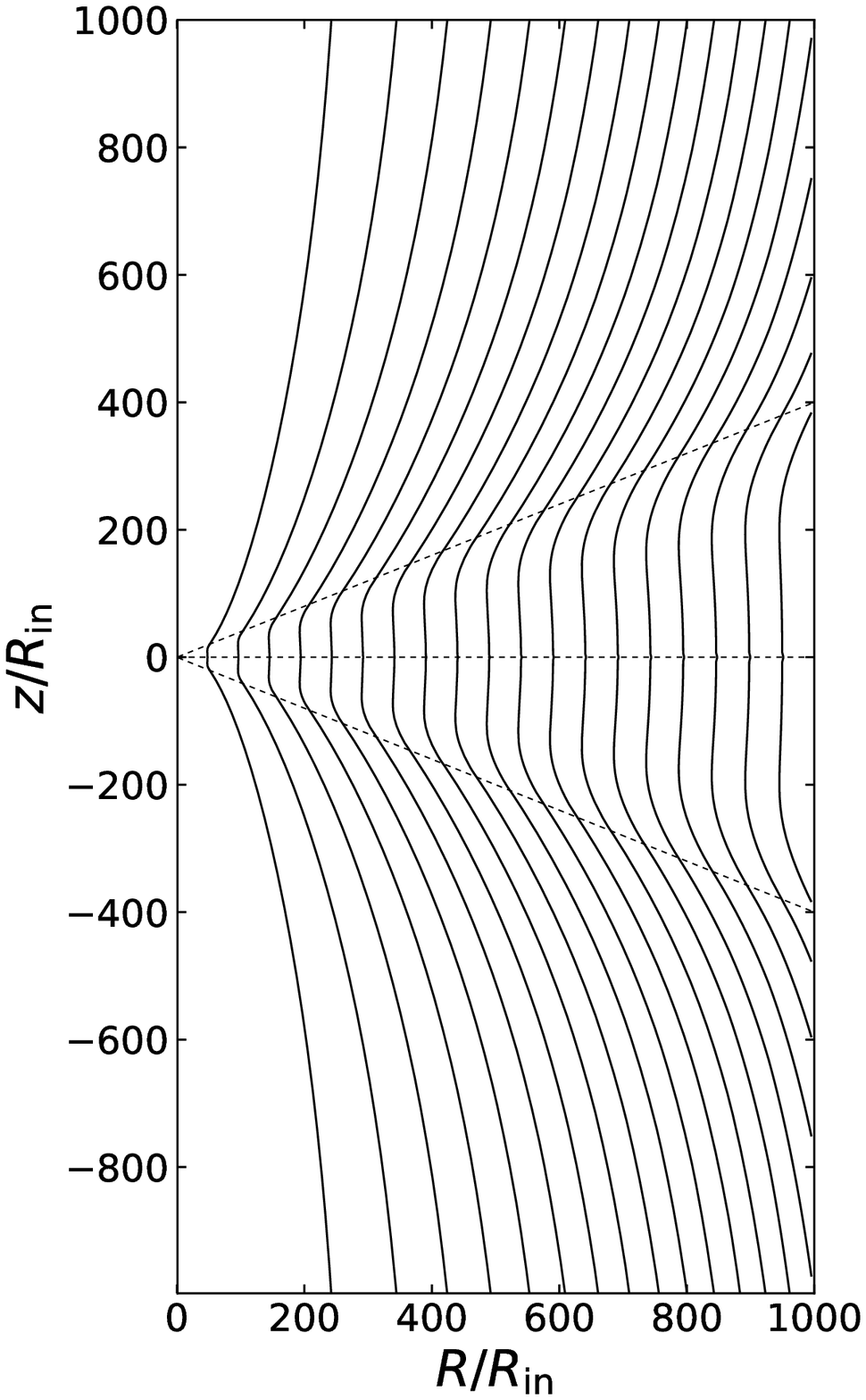}{0.48\textwidth}{}
          }
\centering
\caption{Large-scale poloidal magnetic field configurations. The solid curves show the poloidal magnetic field lines, while the dotted lines show the location of the disk surface (i.e., $z_{\rm h} = 4H$) and the disk mid-plane, respectively. Left panel: $H/R=0.05$. Right panel: $H/R=0.1$.
\label{fig:B_line_og}}
\end{figure}

We calculate the field advection for two different cases. Case 1: We assume the gas to be  isothermal vertically, while the radial velocity is described by changing the values of viscosity parameter as a parameterized function of $z$ to mimic the analytical results of \citet{2012MNRAS.424.2097G,2013MNRAS.430..822G}. {The Prandtl number ${\cal P}_{\rm m,0} = 2.0$ is adopted in all the calculations.} 
We consider the field is advected in a disk with gas extending to a large vertical range, i.e., $|z|\le 4 H$, which is the same as their works. We adopt the relative scale-height $H/R=0.1$ as that used in the numerical simulations in \cite{2018ApJ...857...34Z}, the results of which are in excellent agreement with the analytic results in \cite{2013MNRAS.430..822G}. For comparison, we also calculate two other cases with smaller $H/R=0.05$ and a conventional turbulent thin accretion disk {with $H/R=0.1$ respectively.} The results are shown in Figures \ref{fig:Bz_og} and \ref{fig:B_line_og}. We note that the results are independent of $\alpha$, because the magnetic diffusion is proportional to $\alpha$, and the Prandtl number ${\cal P}_{\rm m,0}=\eta/\nu$ is an input model parameter.    

Case 2: For a disk-corona system, the vertical temperature structure of the disk-corona system is described by Equation \eqref{eq:Theta_z}, in which the parameters $a=3$ and $b=15$ are adopted to mimic a sharp temperature rise in the vertical direction. The corona surface is defined at $z=z_{\rm h}$ by Equation \eqref{eq:epsison}. In the calculations, we adopt the dimensionless temperature $\Theta_0=0.0025$ corresponding to the temperature of a turbulent thin disk with $H/R\sim 0.05$, which is the same as that adopted in the numerical simulations carried out by \cite{2020MNRAS.492.1855M}). A temperature $\Theta _{z_{\rm h}}=0.05$ is adopted in all the calculations (see Equation \ref{eq:Theta_z}). The vertical profiles of the dimensionless temperature are plotted in Figure \ref{fig:DC_Theta}. The gas temperature increases sharply within a narrow region from the main body of the disk to the corona, and the relative gas density is shown in Figure \ref{fig:DC_rho}. With the given disk-corona structure, we calculated the field advection as described in Sections \ref{sec:model} and \ref{sec:num_set}. The magnetic field is efficiently transported to the inner region of the disk with the vertical magnetic field strength increased by several orders of magnitude of the external field strength (see Figure \ref{fig:DC_Bz}). The large-scale magnetic field configurations are plotted in Figure \ref{fig:DC_B_line}. 

\section{discussion}\label{sec:discussion}
The results, calculated in Case 1, are consistent with those obtained in the local 2D analytical study by \cite{2013MNRAS.430..822G}. The magnetic field is efficiently transported inwards by upper fast moving gas layer. Our calculations of the global field advected by such an isothermal disk show that the magnetic field strength in the inner region of the disk increases a lot compared to the external field strength (see the color lines in Figure \ref{fig:Bz_og}), while the field advection is always inefficient in a conventional turbulent thin disk (see the black line in Figure \ref{fig:Bz_og}). Limited by their local analysis, the results obtained by \cite{2013MNRAS.430..822G} on the field advection/diffusion for the same problem is unable to provide any information of the field amplification with radius.   

We note that field advection calculated here is much more significant than the conventional thin disk case with the same value of $H/R$ (see Figure \ref{fig:Bz_og}), because our calculations of the disk extending to $z=4H$,  which means a fast moving gas layer with $z\ga H$ plays a predominant role in the field advection. Such a fast moving gas layer suppresses magnetic diffusion to some extent. We also calculate the large-scale magnetic field configurations. It can be seen in Figure \ref{fig:B_line_og} that the field lines are inclined towards the disk surface significantly in the upper fast moving gas layer. We also compare the results derived with different values of $H/R$, which shows that the field strength amplification for $H/R=0.1$ is about one order of magnitude larger than that for $H/R=0.05$ (Figure \ref{fig:Bz_og}).

\begin{figure}
\centering
\gridline{\fig{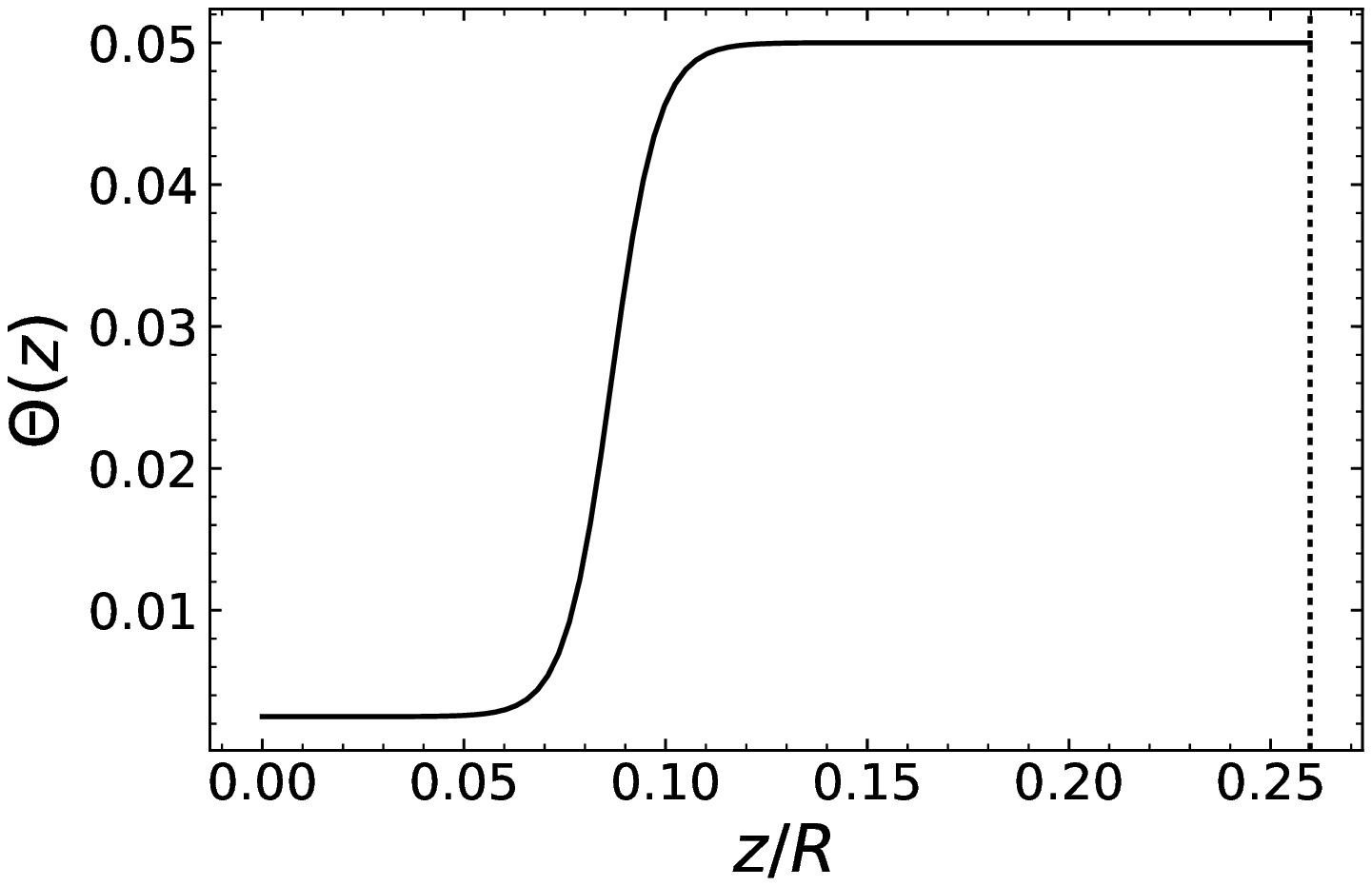}{0.45\textwidth}{}
          \fig{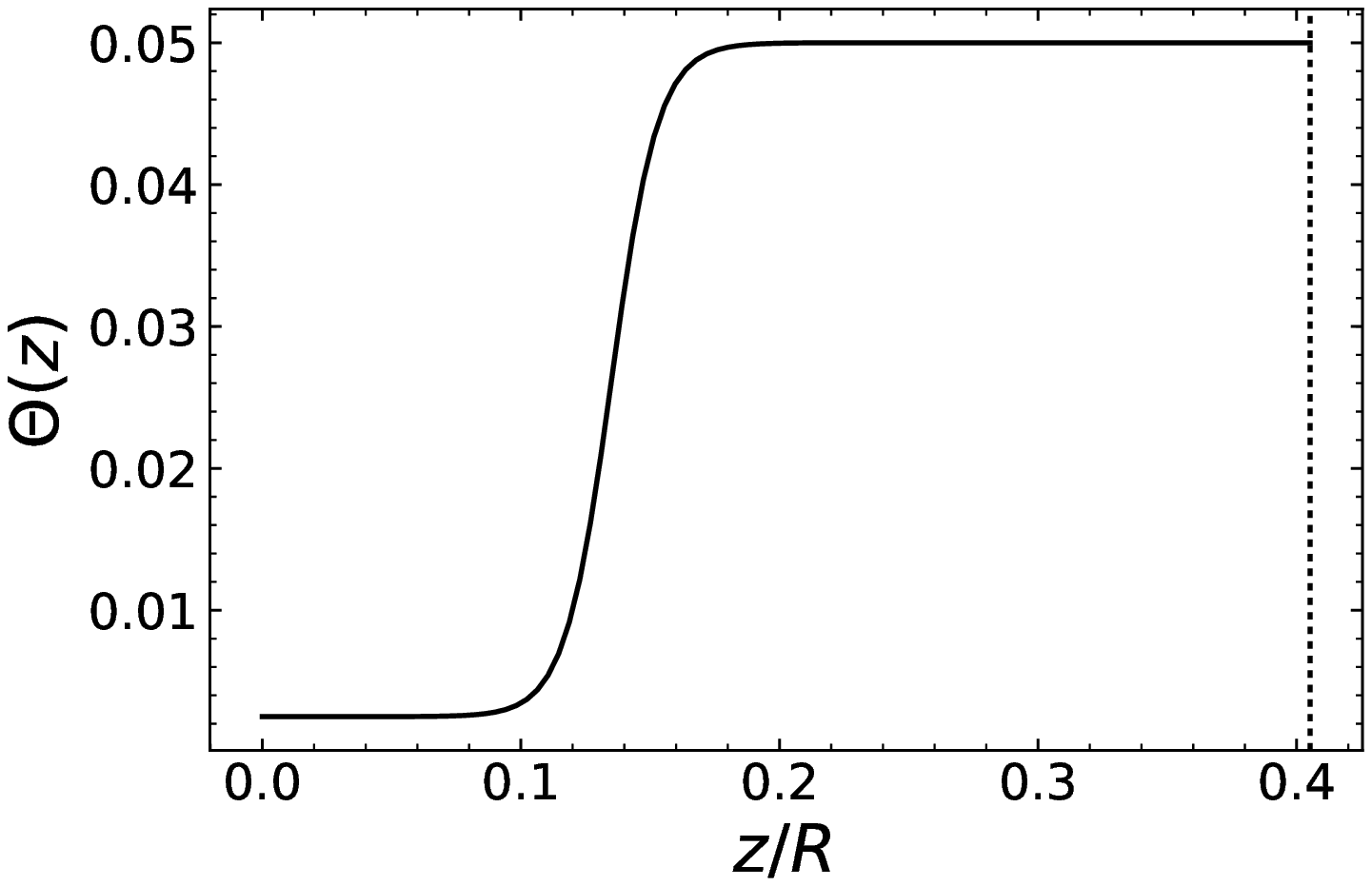}{0.45\textwidth}{}
          }
\centering
\caption{Vertical profiles of the dimensionless gas temperature, the vertical dotted lines are label for the derived location of the disk surface, the parameter $\Theta_0=0.0025$ and $\Theta _{z_{\rm h}}=0.05$ are adopted. Left panel: $\epsilon=10^{-2}$ (see Equation \eqref{eq:epsison}) and the derived location of disk surface is $z_{\rm h}/R = 0.2598$. Right panel: $\epsilon=10^{-3}$ and $z_{\rm h}/R = 0.4052$.
\label{fig:DC_Theta}}
\end{figure}

\begin{figure}
\centering
\gridline{\fig{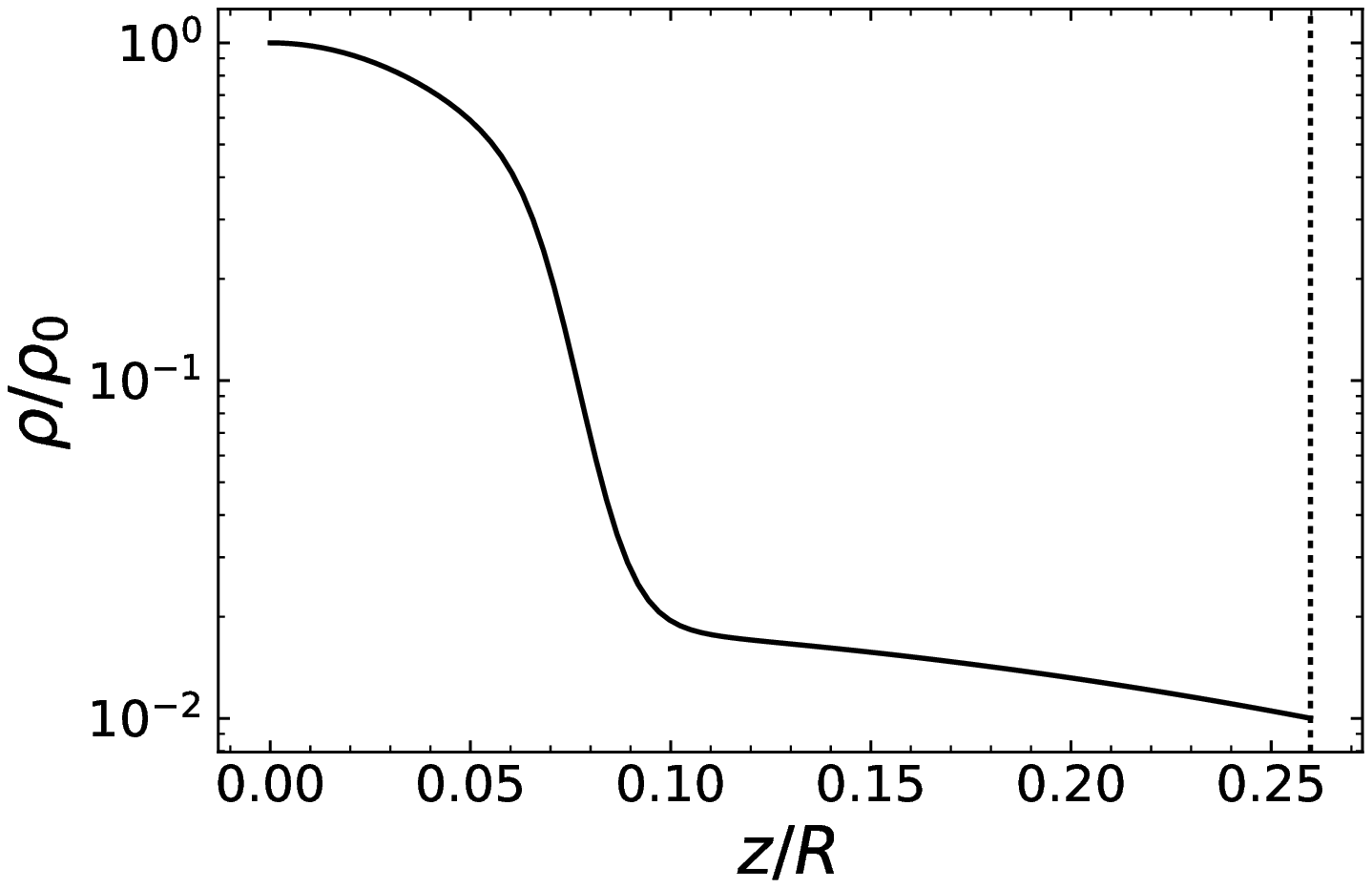}{0.45\textwidth}{}
          \fig{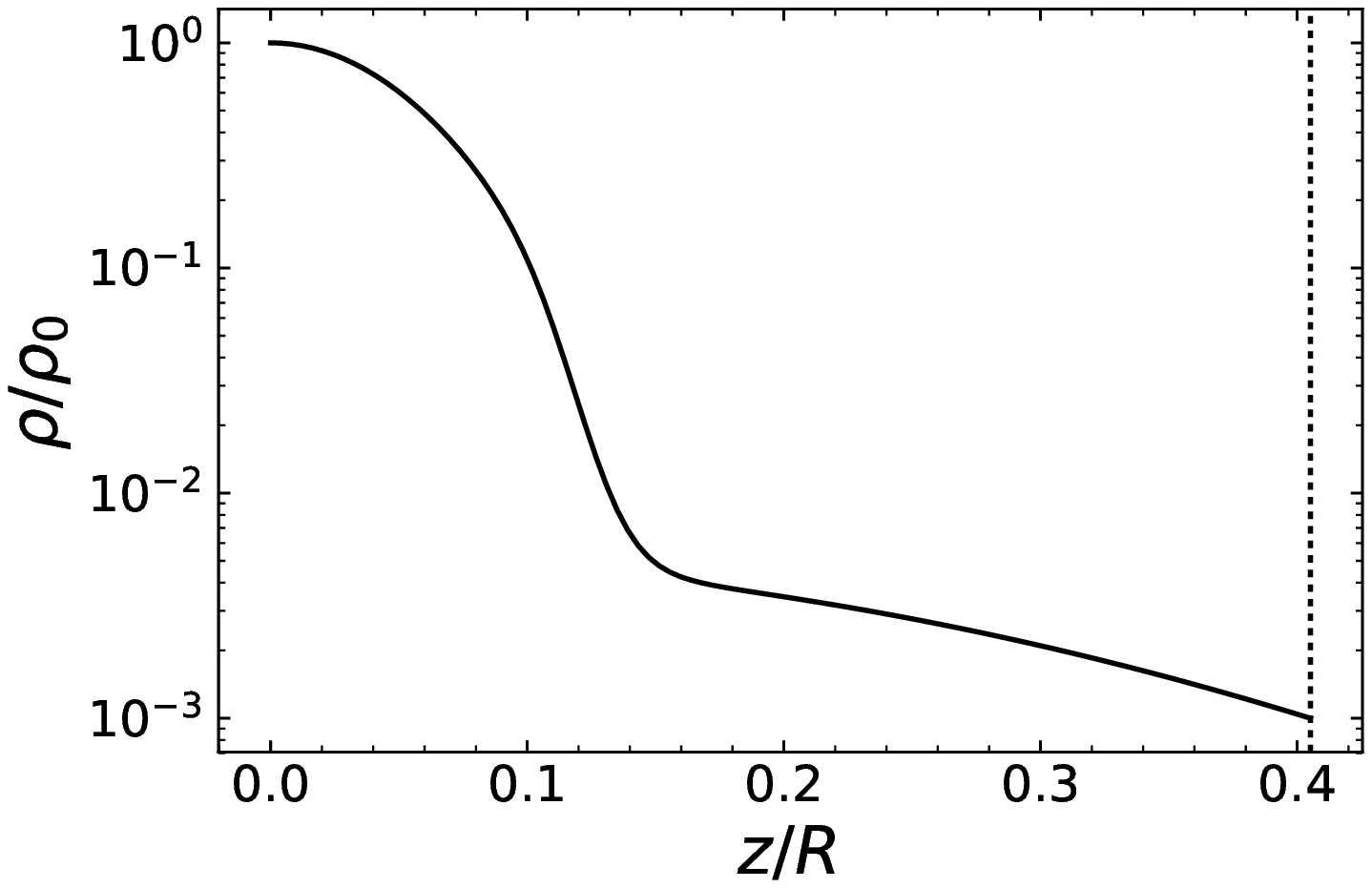}{0.45\textwidth}{}
          }
\centering
\caption{The same as Figure \ref{fig:DC_Theta} but for the vertical profiles of gas density, $\rho_0$ is the gas density at the disk mid-plane.
\label{fig:DC_rho}}
\end{figure}

\begin{figure}
\centering
\includegraphics[width=0.5\columnwidth]{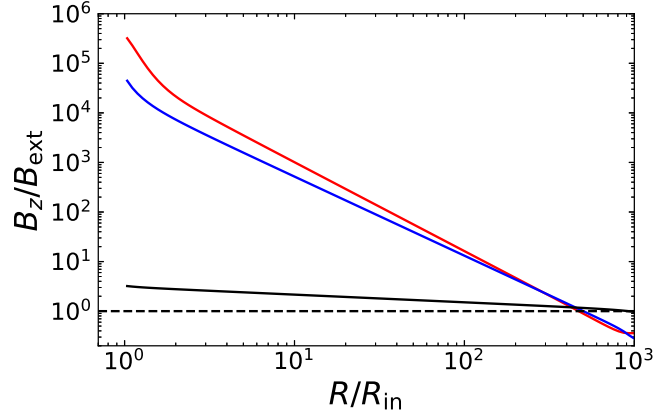}
\caption{The same as Figure \ref{fig:DC_Theta} but for the radial profiles of the vertical component of the poloidal magnetic field. The red and blue lines show the results calculated for $\epsilon = 10^{-3}$ ($z_{\rm h}/R = 0.4052$) and $\epsilon = 10^{-2}$ ($z_{\rm h}/R = 0.2598$), respectively. The black line show the result for a traditional turbulent thin disk with $H/R=0.1$, $B_{\rm ext}$ is the strength of the external imposed magnetic field.   \label{fig:DC_Bz}}
\end{figure}

\begin{figure}
\centering
\gridline{\fig{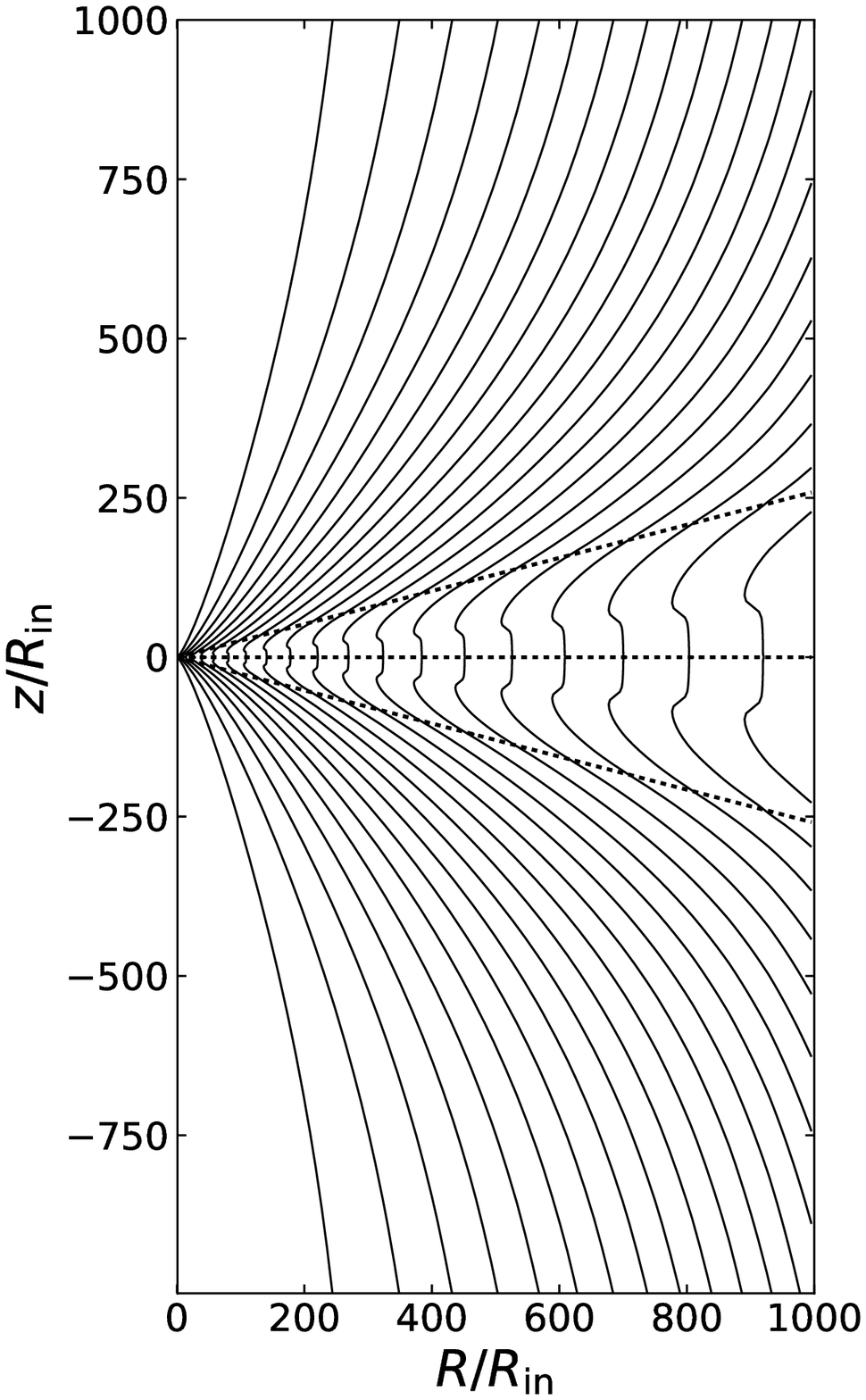}{0.48\textwidth}{}
          \fig{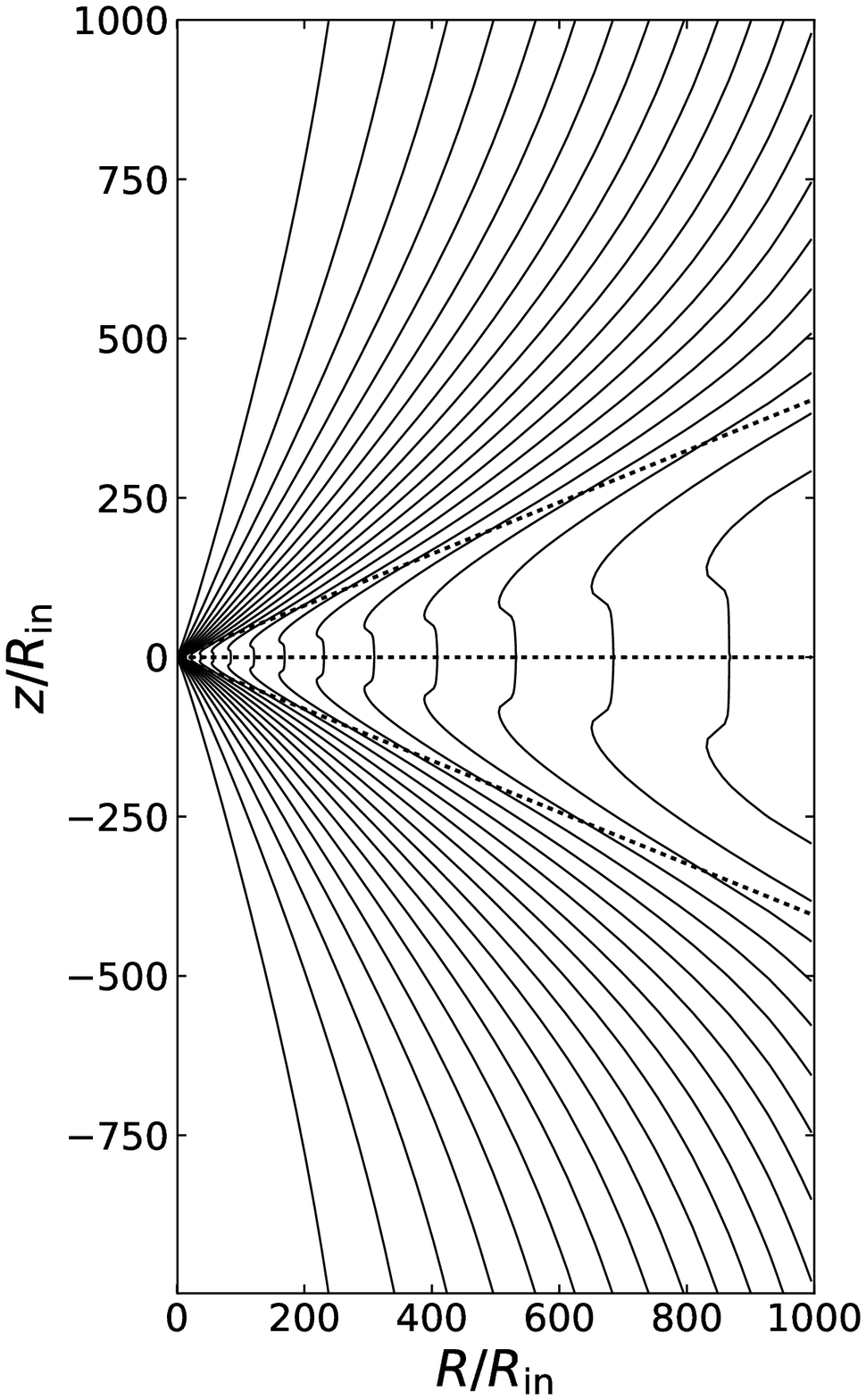}{0.48\textwidth}{}
          }
\centering
\caption{ The same as Figure \ref{fig:DC_Theta} but for the field configurations of the large-scale poloidal magnetic field, dotted lines show the disk surface and mid-plane. Left panel: $\epsilon = 10^{-2}$. Right panel: $\epsilon = 10^{-3}$.
\label{fig:DC_B_line}}
\end{figure}

For the disk-corona system, i.e., Case 2, the temperature increases sharply from $z/R\sim 0.1$ to represent a hot corona (see Figure \ref{fig:DC_Theta}). Given such a vertical temperature profile, the gas density is determined by the vertical hydrostatic equilibrium (see Equation \ref{eq:hydro_equi}), which decreases sharply in the corona to $z=z_{\rm h}$ (see Figure \ref{fig:DC_rho}). In fact, the gas in the upper layer of the corona may be driven by the magnetic field and gas pressure gradient force, which means the corona may connect to an outflow \citep[][]{2005ApJ...635.1203M,2008ApJ...687..156W,2013ApJ...770...31W,2014ApJ...788...71L,2018MNRAS.473.4268C,2019ApJ...872..149L}. The corona can be regarded as a reservoir to feed outflows, however, the transition of the corona to the  outflow is very complicated, and it is sometimes named as outflowing corona \citep*[e.g.,][]{2014ApJ...783..106L}. The calculation of the transition region is beyond the scope of this work. We can still give a rough estimate of the border between the corona and the outflow by the ratio of $p_{\rm m}\sim p_{\rm gas}$, i.e., the gas pressure dominates over the magnetic pressure in the corona, while magnetic pressure is dominant in the outflows. The ratio of gas to magnetic pressure at the disk mid-plane, i.e., $z=0$, is defined as $\beta_{\rm c} = p_{\rm g}(0)/p_{\rm m}(0)$, then we can derive the maximal strength of the field can be advected with respect to the gas pressure at the disk midplane, i.e.,

\begin{equation}\label{eq:beta_min}
    \beta_{\rm c, min} \ga \frac{\rho_{0} c^2_{\rm s, 0}}{\rho_{z_{\rm h}}c^2_{{\rm s},z_{\rm h}}} = \frac{\Theta_0}{\epsilon\Theta_{z_{\rm h}}},
\end{equation}
where $c_{\rm s, 0}$ and $c_{{\rm s},z_{\rm h}}$ are the sound speed at disk mid-plane and the surface of the corona (see Equation \ref{eq:def_Theta}), respectively. Here we have assumed the magnetic field strength at the surface of the corona is close to that at the disk mid-plane. In this work, we use a parameter $\epsilon$ to describe the location of the corona upper surface, and then we can see that, for a small value of $\epsilon$, the maximal strength of the field advected by the hot gas is relatively weak (i.e., a large $\beta_{\rm c}$). For the fixed values of the parameters of $\Theta_0$ and $\Theta_{z_{\rm h}}$, the gas of the upper layer has the same radial velocity, thus a low density gas upper layer can only drag a relatively weak magnetic field inwards, because the gas pressure should be dominant there, hence a large $\beta_{\rm c}$.

The radial velocity of the gas is proportional to the dimensionless gas temperature (see Equations \ref{eq:vR} and \ref{eq:def_Theta}), i.e., the gas in hot corona is moving faster than the cold gas in the region near the midplane \citep[][]{2015ApJ...806..223L,2018ApJ...857...34Z, 2019ApJ...885..144J}. Thus, the magnetic field is efficiently amplified within this fast moving hot gas, and the magnetic field strength can be several orders of magnitude larger than the external field strength (Figure \ref{fig:DC_Bz}). We compare the results calculated with different values of $\epsilon$ in Figure \ref{fig:DC_Bz}. We find that the field amplification for $\epsilon = 10^{-3}$ ($z_{\rm h}/R = 0.4052$) is larger than that for $\epsilon = 10^{-2}$ ($z_{\rm h}/R = 0.2598$). However, one should be cautious that, it does not mean a stronger field is achieved for a lower value of $\epsilon$! A lower $\epsilon$ means a relatively weaker external field the corona can drag inwards, if the values of all the disk parameters are fixed (see Equation \ref{eq:beta_min}). {The disk is relatively thick for a small $\epsilon$, which reduces magnetic diffusion. Thus, the field lines are more inclined towards radial direction (see Figure \ref{fig:DC_B_line}), and the amplification of the field is relatively large than that of a large  $\epsilon$.}

Due to the fast moving corona, the magnetic field is efficiently dragged inwards, and the field lines are  strongly inclined towards the disk plane in the upper layer of the corona (see Figure \ref{fig:DC_B_line}). This field configuration is suitable for launching outflows from the disk \citep[][]{1982MNRAS.199..883B, 1994A&A...287...80C,2013ApJ...765..149C, 2019ApJ...872..149L}. Such field configurations obtained in this work are in excellent agreement with the global MHD simulations of a thin accretion disk with corona by \cite{2018ApJ...857...34Z} and \cite{2020MNRAS.492.1855M}. Powerful outflows are ubiquitously observed in accretion systems with different scales  \citep[e.g.,][]{2009ApJ...701..493R,2010A&A...521A..57T,2013MNRAS.430.1102T,2015Natur.519..436T,2015MNRAS.451.4169G,2017MNRAS.469.1553P,2020ApJ...895...37R}, which may be one of the important feedback mechanisms influencing the host galaxies' dynamics, star formation, or even the growth of their central black holes \citep[][]{2005ApJ...620L..79S,2007ARA&A..45..117M,2012ARA&A..50..455F,2016ApJ...823...90B,2017MNRAS.472..949B,2018ApJ...861..106D,2019ApJ...886...92L,2020ApJ...890...81C}. {Magnetic acceleration could be a main mechanism to driven these powerful outflows \citep[][]{2006Natur.441..953M,2004MNRAS.355.1105F,2015ApJ...814...87M,2015ApJ...805...17F,2018ApJ...853...40F, 2018ApJ...852...35K,2020ApJ...904...30M,2021ApJ...906..105C}.}

We note that the field diffusion is mainly contributed by $\eta \nabla \times \boldsymbol{B}$ (the second term in the right of Equation \ref{eq:induc_B}). It is found that the field lines are almost vertical in the region near the disk midplane (see Figures \ref{fig:B_line_og} and \ref{fig:DC_B_line}), which is caused by the small radial velocity of the gas in that region. This leads to very small $\nabla \times \boldsymbol{B}$ even for a rather strong magnetic field. It means the diffusion in the region near the midplane is substantially suppressed.

In this work, we only consider the advection/diffusion of a poloidal field threading a rotating disk with fast moving gas. It is well known that the radial component of the large-scale magnetic field will be sheared into an azimuthal component due to the differential rotation within the disk, triggering MRI  process \citep[magnetorotational instability:][]{1991ApJ...376..214B}, and the turbulence responsible for angular momentum transport in the disk. In this work, we avoid being involved in such complicated physics, and, instead, we assume that $\alpha$-viscosity can still describe the angular momentum transfer in the disk due to the turbulence triggered by MRI fairly well, as done by many previous works \citep[e.g.,][]{1994MNRAS.267..235L,2013MNRAS.430..822G}. In fact, the results obtained in this work is independent of the value of $\alpha$.

\section{summary}\label{sec:summary}
We study the large-scale magnetic field advection in a turbulent thin accretion disk covered with fast moving gas layer. First, we calculate a global magnetic field configuration of a thin disk with vertical extended isothermal gas, which is roughly a global version of the local analysis by \citet{2013MNRAS.430..822G}.  Their results are confirmed in our calculations, and furthermore we derived the field amplification as a function of the disk radius, which is unavailable in their local analysis. It is found that the field advection in such a thin disk with fast moving gas is much more efficient than that of a conventional thin disk. 

We also explore the field advection in a disk-corona system. The magnetic flux is found to be efficiently transported by the corona, and the field lines are strongly inclined towards the disk surface, which are suitable for launching outflows from the corona. The large-scale magnetic field configurations derived in this work are qualitatively consistent with those obtained in some previous numerical simulations \citep[][]{2018ApJ...857...34Z, 2020MNRAS.492.1855M}. Our results may help understanding the physics of the MHD simulations {and are also useful in explaining the observational features in X-ray binaries and AGNs.}

\acknowledgments

We thank the referee for the valuable comments/suggestions. This work is supported by the NSFC (11773050, 11833007, 12073023), and the CAS grant QYZDJ-SSWSYS023.





\bibliography{ref}
\bibliographystyle{aasjournal}

\end{document}